\documentclass[aps,pre,twocolumn,superscriptaddress,longbibliography]{revtex4-1}
\usepackage{multirow}
\usepackage{epsfig}
\usepackage{xcolor}
\usepackage{graphicx}
\usepackage{dcolumn}
\usepackage{bm}
\usepackage{amsmath}

\begin{document}

\title{An improved belief propagation algorithm for detecting meso-scale structure in complex networks}

\author{Chuang Ma}
\affiliation{School of Mathematical Science, Anhui University, Hefei 230601, China}

\author{Bing-Bing Xiang}
\affiliation{School of Mathematical Science, Anhui University, Hefei 230601, China}

\author{Han-Shuang Chen}
\affiliation{School of Physics and Material Science, Anhui University, Hefei 230601, China}


\author{Hai-Feng Zhang} \email{haifengzhang1978@gmail.com}
\affiliation{School of Mathematical Science, Anhui University, Hefei 230601, China}

\date{\today}

\begin{abstract}

The framework of statistical inference has been successfully used to detect the meso-scale structures in complex networks, such as community structure, core-periphery (CP) structure. The main principle is that the stochastic block model (SBM) is used to fit the observed network and the learnt parameters indicate the group assignment, in which the parameters of model are often calculated via an expectation-maximization (EM) algorithm and a belief propagation (BP) algorithm is implemented to calculate the decomposition itself. In the derivation process of the BP algorithm, some approximations were made by omitting the effects of node's neighbors, the approximations do not hold if networks are dense or some nodes holding large degrees. As a result, for example, the BP algorithm cannot well detect CP structure in networks and even yields wrong detection because the nodal degrees in core group are very large. In doing so, we propose an improved BP algorithm to solve the problem in the original BP algorithm without increasing any computational complexity. By comparing the improved BP algorithm with the original BP algorithm on community detection and CP detection, we find that the two algorithms yield the same performance on the community detection  when the network is sparse, for the community structure in dense networks or CP structure in networks, our improved BP algorithm is much better and more stable. The improved BP algorithm may help us correctly partition different types of meso-scale structures in networks.
\end{abstract}
\maketitle

\section{Introduction} \label{sec:intro}

One important issue in complex networks is the detection of meso-scale structures, which has received many attentions from a variety of scientific disciplines, such as, community detection and CP structure detection. Community detection aims to partition the nodes in a network into groups such that
the edges within community are densely connected, but the edges bridging different communities are sparse. The study on the community detection is a hot topic and many algorithms have been developed~\cite{fortunato2010community}, e.g., the algorithms based on modularity~\cite{newman2006modularity}, spectral clustering~\cite{white2005spectral}, hierarchical clustering~\cite{lancichinetti2009detecting,girvan2002community}, nonnegative matrix factorization approach~\cite{yang2013overlapping}, clique percolation theory~\cite{palla2005uncovering}, and so on. Recently,  another type of meso-scale structure---CP structure has also attracted some attentions because such a meso-scale structure is different from community structure and commonly exists in social networks, transportation networks as well as biological networks~\cite{white1976social,borgatti2000models,rombach2017core,verma2016emergence,chen2018double}.
The purpose of the CP detection is to partition the nodes in a network into groups such that nodes in core group are more connected both to other core nodes and to peripheral nodes, but nodes in peripheral group are less connected to each other~\cite{kojaku2017finding,xiang2018unified,della2013profiling,ma2018detection,holme2005core,kojaku2018core}.

Both the community structure and CP structure can be uniformly viewed as the block structure. One effective approach in discovering block structure is the statistical inference framework, in which a generative model-SBM is adopted to fit the network data and learns the parameters of the model. The learnt parameters in SBM can discover the block structures (i.e., group assignment), including community structure and CP structure~\cite{decelle2011inference,decelle2011asymptotic,zhang2014scalable,karrer2011stochastic}. The parameters in SBM are often solved by using algorithms such as Monte Carlo (MC) sampling and EM algorithm, meanwhile,  the marginal probabilities in the M-step of EM algorithm can be calculated by the BP algorithm, which is a useful tool to approximately solve the problem of statistical inference~\cite{decelle2011inference,zhang2014scalable}. One assumption in the BP algorithm is that the neighbors' effect can be ignored, leading to one quantity in iterative formula becomes an external field and is independent of the special nodes ~\cite{decelle2011asymptotic,zhang2015identification}. Because nodes in core group often connect to many nodes and their neighbors' effect is huge and cannot be ignored, as a result, the approximation may yield big error when the BP algorithm is used to detect CP structure. Therefore, in this paper, we modify in the derivation of the BP algorithm by circumventing the assumption. The improved BP algorithm does not increase any computational complexity, but a more accurate iterative formula is obtained. By using the improved BP algorithm to detect community structure and CP structure, we find that the two BP algorithms have the same precision in the community detection when networks are sparse. As for the CP structure and the community structure in dense networks, our improved BP algorithm is more accurate and more stable. It is important to note that the original BP algorithm becomes completely invalid when the CP structure is dominant (i.e., nodes in core groups connect to most of nodes in network), on the contrary, the improved BP algorithm can well detect CP structure.

\section{Statistical inference} \label{sec:method}
Given an undirected network, our goal is to learn the parameters in the SBM model by best fitting the network data. The problem of maximum likelihood estimation can be solved by implementing the EM algorithm.

\subsection{EM algorithm}\label{EM}

Let $\bm{A}$ be the adjacency matrix of the network, where elements $A_{ij}=A_{ji}=1$ if nodes $i$ and $j$ are connected, otherwise,  $A_{ij}=A_{ji}=0$ . There are several parameters in the SBM, including $k$ (the number of groups), an $k\times k$ \emph{affinity matrix} $p$ (element $p_{rs}$ denotes the probability of an edge between group
$r$ and group $s$), and $\gamma$ (an array that
determines the relative size of each group). Our goal is to maximize the probability, or likelihood $P(\bm{A}|p,\gamma)$, which includes a group assignment $\{g_i\}$ (the group to which node $i$ belongs). So we have
\begin{eqnarray}\label{likelihood}
P\left( {\bm{A}|p,\gamma } \right) = \sum\limits_g {P\left( {\bm{A},g|p,\gamma } \right)},
\end{eqnarray}
where
\begin{eqnarray}\label{condi_pro}
\nonumber P\left( {\bm{A},g|p,\gamma } \right) &=& P\left( {\bm{A}|g,p,\gamma } \right)P\left( {g|\gamma } \right) \\&=& \prod\limits_{i < j} {p_{{g_i}{g_j}}^{{A_{ij}}}{{\left( {1 - {p_{{g_i}{g_j}}}} \right)}^{1 - {A_{ij}}}}} \prod\limits_i {{\gamma _{{g_i}}}}.
\end{eqnarray}

To determine the most likely values of the parameters $p$ and $\gamma$, the common way is to maximize their logarithm of the
likelihood  with respect to them. But it is still very difficult to solve the values of the $p$ and $\gamma$ by maximizing the logarithm of the
likelihood. This problem can be solved by well-known EM method in statistics~\cite{EM1977}. At first, according to Jensen's inequality, one has
\begin{eqnarray}\label{Jensen}
\nonumber\log \sum\limits_g {P\left( {\bm{A},g|p,\gamma } \right)} & \ge& \sum\limits_g {q\left( g \right)\log \frac{{P\left( {\bm{A},g|p,\gamma } \right)}}{{q\left( g \right)}}}\\& \overset{\Delta} = &L\left( {p,\gamma } \right),
\end{eqnarray}
where
\begin{eqnarray}\label{condition}
\nonumber q\left( g \right) &=& \frac{{P\left( {\bm{A},g|p,\gamma } \right)}}{{\sum\limits_g {P\left( {A,g|p,\gamma } \right)} }}\\& =& \frac{{\prod\limits_{i < j} {p_{{g_i}{g_j}}^{{A_{ij}}}{{\left( {1 - {p_{{g_i}{g_j}}}} \right)}^{1 - {A_{ij}}}}} \prod\limits_i {{\gamma _{{g_i}}}} }}{{\sum\limits_g {\prod\limits_{i < j} {p_{{g_i}{g_j}}^{{A_{ij}}}{{\left( {1 - {p_{{g_i}{g_j}}}} \right)}^{1 - {A_{ij}}}}} \prod\limits_i {{\gamma _{{g_i}}}} } }}.
\end{eqnarray}
In fact $q\left( g \right) = P\left( {g|\bm{A},p,\gamma } \right)$ is the probability distribution with respect to the group assignment $g$.

Therefore, maximizing $\log P\left( {\bm{A}|p,\gamma } \right)$ with
respect to $p$ and $\gamma$ can be done as follows: first, choosing $q(g)$ as in Eq.~(\ref{condition}) to make the two sides of Eq.~(\ref{Jensen}) equal, then maximizing $L(p,\gamma)$ with respect to the parameters. We further have
\begin{widetext}
\begin{eqnarray}\label{simplicity}
\begin{array}{l}
L\left( {p,\gamma } \right) = \sum\limits_g {q\left( g \right)\log \frac{{P\left( {\bm{A},g|p,\gamma } \right)}}{{q\left( g \right)}}} \\
 = \sum\limits_g {q\left( g \right)\log \left[ {\prod\limits_{i < j} {p_{{g_i}{g_j}}^{{A_{ij}}}{{\left( {1 - {p_{{g_i}{g_j}}}} \right)}^{1 - {A_{ij}}}}} \prod\limits_i {{\gamma _{{g_i}}}} } \right]}  - \sum\limits_g {q\left( g \right)\log q\left( g \right)} \\
 = \sum\limits_g {q\left( g \right)\left[ {\sum\limits_{i < j} {\left[ {{A_{ij}}\log {p_{{g_i}{g_j}}} + \left( {1 - {A_{ij}}} \right)\log \left( {1 - {p_{{g_i}{g_j}}}} \right)} \right]}  + \sum\limits_i {\log {\gamma _{{g_i}}}} } \right]}  - \sum\limits_g {q\left( g \right)\log q\left( g \right)} \\
 = \frac{1}{2}\sum\limits_{i \ne j} {\sum\limits_{rs} {\left[ {{A_{ij}}q_{rs}^{ij}\log {p_{rs}} + \left( {1 - {A_{ij}}} \right)q_{rs}^{ij}\log \left( {1 - {p_{rs}}} \right)} \right]} }  + \sum\limits_{ir} {q_r^i\log {\gamma _r}}  - \sum\limits_g {q\left( g \right)\log q\left( g \right)},
\end{array}
\end{eqnarray}
\end{widetext}
where $q^{i}_r$ is the marginal probability within the joint probability
distribution $q(g)$ that node $i$ belongs to group $r$, namely,
\begin{eqnarray}\label{one_marginal}
q_r^i = \sum\limits_g {q\left( g \right){\delta _{{g_i},r}}}.
\end{eqnarray}
Similarly,  $q^{ij}_{rs}$ is the probability node $i$ and node $j$ belonging to group $r$ and group $s$, respectively, which is expressed as:
\begin{eqnarray}\label{two_marginal}
q_{rs}^{ij} = \sum\limits_g {q\left( g \right){\delta _{{g_i},r}}{\delta _{{g_j},s}}}.
\end{eqnarray}

Within the condition $\sum\limits_r {{\gamma _r}}=1$, taking the partial derivative of Eq.~(\ref{simplicity}) with respect to $p$ and $\gamma$, and setting them to be zero, leading to
\begin{eqnarray}\label{affine}
{p_{rs}} = \frac{{\sum\limits_{i \ne j} {{A_{ij}}q_{rs}^{ij}} }}{{\sum\limits_{i \ne j} {q_{rs}^{ij}} }}
\end{eqnarray}
and
\begin{eqnarray}\label{gamma}
{\gamma _r} = \frac{1}{n}\sum\limits_i {q_r^i}.
\end{eqnarray}
Thus we can calculate the two parameters in SBM by numerical iterations in EM algorithm. Given the initial conditions $p$ and $\gamma$, the distribution $g(q)$ can be first obtained from Eq.~(\ref{condition}), then $q^i_r$ and $q^{ij}_{rs}$ are solved from Eqs.~(\ref{one_marginal}) and (\ref{two_marginal}), the new estimations of $p$ and $\gamma$ are computed based on Eqs. (\ref{affine}) and (\ref{gamma}). Repeating the above process until it converges to a local maximum of the log-likelihood. In general, we implement the above iteration process several times with different initial conditions to yield the global maximum of the log-likelihood.

All possible pairs should be considered when calculating the denominator of Eq.~(\ref{affine}), the computational complexity is high. We can simplify Eq.~(\ref{affine}) based on the following mean-field approximation:
\begin{eqnarray}\label{meanfield}
\nonumber\sum\limits_{ij} {q_{rs}^{ij}}  &=& \sum\limits_g {q\left( g \right)\sum\limits_i {{\delta _{{g_i},r}}} \sum\limits_j {{\delta _{{g_j},s}}} }  = \sum\limits_g {q\left( g \right){n_r}{n_s}} \\& =& \left\langle {{n_r}{n_s}} \right\rangle  \simeq \left\langle {{n_r}} \right\rangle \left\langle {{n_s}} \right\rangle,
\end{eqnarray}
where $\langle\cdots\rangle$ denotes the expectation within the probability distribution $q(g)$, thus,
\begin{eqnarray}\label{number}
\left\langle {{n_r}} \right\rangle  = \sum\limits_g {q\left( g \right)\sum\limits_i {{\delta _{{g_i},r}}} }  = \sum\limits_i {q_r^i}.
\end{eqnarray}
In this situation, equation~(\ref{affine}) can be simply expressed as
\begin{eqnarray}\label{simplified}
{p_{rs}} = \frac{{\sum\limits_{i \ne j} {{A_{ij}}q_{rs}^{ij}} }}{{\sum\limits_i {q_r^i} \sum\limits_j {q_s^j} }}.
\end{eqnarray}
One can see the denominator in Eq.~(\ref{simplified}) is easier to be calculated than that in Eq.~(\ref{affine}) since we only need to sum over all nodes but not to all possible pairs.

The obtained type of structure usually depends on the selection of the initial values. The key part of the EM algorithm is the calculation of Eq.~(\ref{condition}), which can be calculated by MC sampling. Because the sampling space is large (i.e, the sampling space is $k^n$ if there are $k$ groups). The results are unstable when the network is large. The BP algorithm can well solve this problem, moreover,  which can deal with networks with large sizes.

\subsection{original BP algorithm and improved BP algorithm}
The BP algorithm is a message passing technique to solve the marginal probability of the probability distribution $q(g)$.  The message $\eta^{i\to j}$ is defined as the probability that node $i$ belongs to group
$r$ when node $j$ is removed from the network (cavity theory in statistical physics)~\cite{lamperti2012stochastic}, which is read as
\begin{eqnarray}\label{cavity}
&&\eta _r^{i \to j} =\\&&
 \nonumber\frac{{{\gamma _r}}}{{{Z_{i \to j}}}}\prod\limits_{k \in V/{N^ * }\left( i \right)} {\sum\limits_s {\eta _s^{k \to i}\left( {1 - {p_{rs}}} \right)} } \prod\limits_{k \in N\left( i \right)/j} {\sum\limits_s {\eta _s^{k \to i}{p_{rs}}} },
 \end{eqnarray}
where $N(i)$ is the neighborhood set of node $i$, and $N^*(i)=N(i)\cup {i}$. ${Z_{i \to j}}$ is a normalizing constant.

When the node $j$ is not the neighbor of node $i$ and when the network is large, $\eta _r^{i \to j}$ approximated by $q^i_r$. It can be understood that the removal of the node $j$ has no influence on the group assignment of node $i$ if node $j$ is not its neighbor. Thus, when the network is large and sparse, it can be approximated as
\begin{eqnarray}\label{approximation}
\begin{array}{l}
\prod\limits_{k \in V/{N^ * }\left( i \right)} {\sum\limits_s {\eta _s^{k \to i}\left( {1 - {p_{rs}}} \right)} } \approx \prod\limits_{k \in V/{N^ * }\left( i \right)} {\sum\limits_s {q_s^k\left( {1 - {p_{rs}}} \right)} } \\
 \approx \prod\limits_{k \in V} {\sum\limits_s {q_s^k\left( {1 - {p_{rs}}} \right)} }  = \prod\limits_{k \in V} {\left[ {1 - \sum\limits_s {q_s^k{p_{rs}}} } \right]}.
\end{array}
 \end{eqnarray}

\emph{Remark 1:} the approximation in Eq.~(\ref{approximation}) indicates that the size of the set $N^*(i)$ is negligible when the network is large and sparse, Namely, $V/N^*(i)\approx V$. In this situation, the quantity  in Eq.~(\ref{approximation}) is an external field and is independent of one special node. We will demonstrate such an approximation is unreasonable for the detection of CP structures or when the networks are dense, and the approximation may yield fatal mistakes under some situations.

According to Eq.~(\ref{approximation}), equation~(\ref{cavity}) is rewritten as
\begin{eqnarray}\label{message}
&&\eta _r^{i \to j} = \\
\nonumber&&\frac{{{\gamma _r}}}{{{Z_{i \to j}}}}\prod\limits_k {\left[ {1 - \sum\limits_s {q_s^k{p_{rs}}} } \right]} \prod\limits_{k \in N\left( i \right)/j} {\sum\limits_s {\eta _s^{k \to i}{p_{rs}}} }.
 \end{eqnarray}
In Eq.~(\ref{message}), normalizing constant $Z_{i \to j}$ is
 \begin{eqnarray}\label{normalize}
&&{Z_{i \to j}} =\\
 &&\nonumber\sum\limits_r {{\gamma _r}\prod\limits_k {\left[ {1 - \sum\limits_s {q_s^k{p_{rs}}} } \right]} \prod\limits_{k \in N\left( i \right)/j} {\sum\limits_s {\eta _s^{k \to i}{p_{rs}}} } }.
 \end{eqnarray}

From the iteration formula in Eq.~(\ref{message}), the marginal distribution $q^i_r$ is calculated as
 \begin{eqnarray}\label{q_i_r}
 q_r^i = \frac{{{\gamma _r}}}{{{Z_i}}}\prod\limits_k {\left[ {1 - \sum\limits_s {q_s^k{p_{rs}}} } \right]} \prod\limits_{k \in N\left( i \right)} {\sum\limits_s {\eta _s^{k \to i}{p_{rs}}} },
  \end{eqnarray}
where
 \begin{eqnarray}\label{z_i}
 {Z_i} = \sum\limits_r {{\gamma _r}\prod\limits_k {\left[ {1 - \sum\limits_s {q_s^k{p_{rs}}} } \right]} \prod\limits_{k \in N\left( i \right)} {\sum\limits_s {\eta _s^{k \to i}{p_{rs}}} } } .
   \end{eqnarray}
At the same time, the probability distribution $q_{rs}^{ij}$ is described as
\begin{eqnarray}\label{q_ij_rs}
 q_{rs}^{ij} = \frac{{\eta _r^{i \to j}\eta _s^{j \to i}{p_{rs}}}}{{\sum\limits_{rs} {\eta _r^{i \to j}\eta _s^{j \to i}{p_{rs}}} }}.
\end{eqnarray}

In sum, the BP algorithm is a part of M-step in EM algorithm, the main steps of EM algorithm are summarized in Algorithm 1.\\
\textbf{Algorithm 1:}

\textbf{Step 1:} Initialize the parameters $p$ and $\gamma$;

\textbf{Step 2:} Initialize the parameters $\eta$ and $q$;

\textbf{Step 3:} Iterate Eqs.~(\ref{message}) and (\ref{q_i_r}) until convergence, and the marginal probability of each node $q^i_r$ is obtained. Meanwhile, the marginal probability $q^{ij}_{rs}$ is calculated according to Eq.~(\ref{q_ij_rs});

\textbf{Step 4:} New estimations of $p$ and $\gamma$ are computed from Eqs.~(\ref{gamma}) and (\ref{simplified});

\textbf{Step 5:} Repeat step 2--step 4 until the parameters converge and each node $i$ is assigned to the group $r$ with the highest probability $q^i_r$.

As we have mentioned in Remark 1, equation~(\ref{approximation}) indicates that the set $N^*(i)$ is negligible and the quantity becomes an external field. Indeed, the approximation is not suitable for the CP structure or the networks with dense connections. For example, the nodes in core group are more likely to connect to most of nodes, i.e., the size of set $N^*(i)\approx V$ for core node $i$. In this case, the approximation in Eq.~(\ref{approximation}) (i.e, $\prod\limits_{k \in V/{N^ * }\left( i \right)} {\sum\limits_s {q_s^k\left( {1 - {p_{rs}}} \right)} }$  $\approx$ $\prod\limits_{k \in V} {\sum\limits_s {q_s^k\left( {1 - {p_{rs}}} \right)} }$ ) is evidently unreasonable. To avoid the mistakes induced from the approximation in Eq.~(\ref{approximation}), one can modify Eq.~(\ref{condition}) in a new form

 \begin{eqnarray}\label{new_q_g}
 \nonumber q\left( g \right) &=& \frac{{\prod\limits_{i < j} {p_{{g_i}{g_j}}^{{A_{ij}}}{{\left( {1 - {p_{{g_i}{g_j}}}} \right)}^{1 - {A_{ij}}}}} \prod\limits_i {{\gamma _{{g_i}}}} }}{{\sum\limits_g {\prod\limits_{i < j} {p_{{g_i}{g_j}}^{{A_{ij}}}{{\left( {1 - {p_{{g_i}{g_j}}}} \right)}^{1 - {A_{ij}}}}} \prod\limits_i {{\gamma _{{g_i}}}} } }}
  \\&=& \frac{{\prod\limits_{i < j} {{{\left( {\frac{{{p_{{g_i}{g_j}}}}}{{1 - {p_{{g_i}{g_j}}}}}} \right)}^{{A_{ij}}}}\left( {1 - {p_{{g_i}{g_j}}}} \right)} \prod\limits_i {{\gamma _{{g_i}}}} }}{{\sum\limits_g {\prod\limits_{i < j} {{{\left( {\frac{{{p_{{g_i}{g_j}}}}}{{1 - {p_{{g_i}{g_j}}}}}} \right)}^{{A_{ij}}}}\left( {1 - {p_{{g_i}{g_j}}}} \right)} \prod\limits_i {{\gamma _{{g_i}}}} } }}.
 \end{eqnarray}
 According to Refs.~\cite{zhang2014scalable,shi2018weighted}, the message $\eta^{i\to j}_r$ can be rewritten as
\begin{eqnarray}\label{new_eta}
&& \eta _r^{i \to j} =\\&&\nonumber \frac{{{\gamma _r}}}{{{Z_{i \to j}}}}\prod\limits_{k \in V/\left\{ {i,j} \right\}} {\sum\limits_s {\eta _s^{k \to i}\left( {1 - {p_{rs}}} \right)} } \prod\limits_{k \in N\left( i \right)/j} {\sum\limits_s {\eta _s^{k \to i}\frac{{{p_{rs}}}}{{1 - {p_{rs}}}}} },
  \end{eqnarray}
where
\begin{eqnarray}\label{new_approximation}
&&\nonumber \prod\limits_{k \in V/\left\{ {i,j} \right\}} {\sum\limits_s {\eta _s^{k \to i}\left( {1 - {p_{rs}}} \right)} } \simeq \prod\limits_{k \in V/\left\{ {i,j} \right\}} {\left[ {1 - \sum\limits_s {q_s^k{p_{rs}}} } \right]}  \\&&\simeq \prod\limits_{k \in V} {\left[ {1 - \sum\limits_s {q_s^k{p_{rs}}} } \right]}.
 \end{eqnarray}

\emph{Remark 2:} The approximation in Eq.~(\ref{new_approximation}) is negligible because $\eta^{k\to i}_s$ approximately equals to $q^k_s$ when $k \in V/\left\{ {i,j} \right\}$ and only one edge is removed (i.e., $V/\{i,j\}\simeq V$).

Now, the iteration in BP algorithm is described as:
\begin{eqnarray}\label{new_message}
&&\eta _r^{i \to j} =\\&&\nonumber \frac{{{\gamma _r}}}{{{Z_{i \to j}}}}\prod\limits_k {\left[ {1 - \sum\limits_s {q_s^k{p_{rs}}} } \right]} \prod\limits_{k \in N\left( i \right)/j} {\sum\limits_s {\eta _s^{k \to i}\frac{{{p_{rs}}}}{{1 - {p_{rs}}}}} },
 \end{eqnarray}
where $Z_{i\to j}$ is
\begin{eqnarray}\label{new_normalize}
&&{Z_{i \to j}} =\\&&\nonumber\sum\limits_r {{\gamma _r}\prod\limits_k {\left[ {1 - \sum\limits_s {q_s^k{p_{rs}}} } \right]} \prod\limits_{k \in N\left( i \right)/j} {\sum\limits_s {\eta _s^{k \to i}\frac{{{p_{rs}}}}{{1 - {p_{rs}}}}} } }.
 \end{eqnarray}
The marginal probability $q^i_r$ is given as
\begin{eqnarray}\label{new_q_i_r}
&&q_r^i = \\&&\nonumber\frac{{{\gamma _r}}}{{{Z_i}}}\prod\limits_k {\left[ {1 - \sum\limits_s {q_s^k{p_{rs}}} } \right]} \prod\limits_{k \in N\left( i \right)} {\sum\limits_s {\eta _s^{k \to i}\frac{{{p_{rs}}}}{{1 - {p_{rs}}}}} },
 \end{eqnarray}
 where
\begin{eqnarray}\label{new_z_i}
&&{Z_i} = \\&&\nonumber\sum\limits_r {{\gamma _r}\prod\limits_k {\left[ {1 - \sum\limits_s {q_s^k{p_{rs}}} } \right]} \prod\limits_{k \in N\left( i \right)} {\sum\limits_s {\eta _s^{k \to i}\frac{{{p_{rs}}}}{{1 - {p_{rs}}}}} } } .
\end{eqnarray}
Also, the probability $q^{ij}_{rs}$ satisfies
\begin{eqnarray}\label{new_q_ij_rs}
q_{rs}^{ij} = \frac{{\eta _r^{i \to j}\eta _s^{j \to i}\frac{{{p_{rs}}}}{{1 - {p_{rs}}}}}}{{\sum\limits_{rs} {\eta _r^{i \to j}\eta _s^{j \to i}\frac{{{p_{rs}}}}{{1 - {p_{rs}}}}} }}.
\end{eqnarray}

Compared with Eqs.~(\ref{message}-\ref{z_i}), the only modification is that $p_{rs}$ is replaced by $\frac{{{p_{rs}}}}{{1 - {p_{rs}}}}$, which does not add any extra computational complexity, however, our method is more suitable for networks with large nodes, such as CP structure. Next we will verify the advantages of our improved BP algorithm.

\section{Detection of meso-scale structures}
In this section, to measure the accuracy of different algorithms on the detection of meso-scale structures in synthetic networks, the normalized mutual information (NMI) index is used, which is defined as~\cite{pizzuti2012multiobjective}:
 \begin{eqnarray}\label{NMI}
NMI(X, Y)=\frac{2I(X,Y)}{H(X)+H(U)}.
\end{eqnarray}
Here $X$ and $Y$ are the partition determined by algorithms and the real partition, respectively, $I(X,Y)$ is the mutual information of $X$ and $Y$. $H(X)$ and $H(Y)$ are the entropy of $X$ and $Y$, respectively.

\subsection{community detection}
We apply the SBM to generate synthetic networks with community structures to compare the performance of the improved and original BP algorithms.  Assuming that the number of communities is $k=2$, and the size of each community is the same. As a result, the probability of connection between the communities is $p_{rs}=c_{out}/n$ if $r\neq s$ and $p_{rs}=c_{in}/n$  if $r=s$ ($n$ is the network size). We use $\varepsilon  = {c_{out}}/{c_{in}}$ to denote the ratio between these two entries. For the community structures, $\varepsilon<1$. Smaller value of $\varepsilon$ gives rise to the stronger community structure. For a given average degree $c = \left( {{c_{in}} + q\left( {{c_{out}} - 1} \right)} \right)/q$, there is a critical value which determines whether the community structure is detectable~\cite{zhang2014scalable}
\begin{eqnarray}\label{threshold}
{\varepsilon ^ * } = \frac{{\sqrt c  - 1}}{{\sqrt c  - 1 + k}}.
\end{eqnarray}
When $\varepsilon  < {\varepsilon ^ * }$, BP algorithm can detect the community structure, otherwise, neither BP algorithm nor other algorithms can detect the community structure when $\varepsilon  > {\varepsilon ^ * }$.

\begin{figure}
\includegraphics[width=\linewidth]{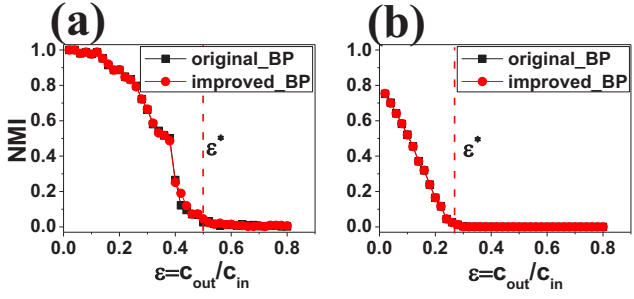}
\caption{(Color online) The comparison of improved and original BP algorithms on the community detection for the sparse networks. (a) $c=9$, $n=200$ and $\varepsilon ^ * =0.5$. (b) $c=3$, $n=10000$ and $\varepsilon ^ * =0.27$. Red dashed line indicates the threshold value $\varepsilon ^ *$. The results are obtained by ensemble averaging over 10 independent realizations.}
\label{fig1}
\end{figure}

We generate two synthetic networks with small average degree to compare the performance of the two algorithms. From Fig.~\ref{fig1}, one can see that the performance of the two algorithms are essentially the same. It is because that all nodal degrees in the SBM are small, the approximation in Eq.~(\ref{approximation}) is reasonable.

\begin{figure}
\includegraphics[width=\linewidth]{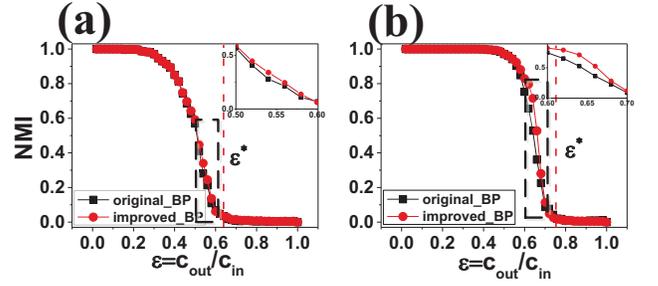}
\caption{(Color online) The comparison of improved and original BP algorithms on the community detection for the dense networks. (a)  $c=20$ and $\varepsilon ^ * =0.6345$. (b) $c=50$ and $\varepsilon ^ * =0.7522$. The network size $n=200$ for the two networks. Red dashed line indicates the threshold value $\varepsilon ^ *$. The inset is the enlarged figure of the region marked by frame. The results are obtained by ensemble averaging over 100 independent realizations.}
\label{fig1new}
\end{figure}

We then compare the two algorithms in synthetic networks with larger average degrees. As shown in Fig.~\ref{fig1new}, the performance of the improved BP algorithm is better than that of the original BP algorithm when $\varepsilon$ is slightly larger than the threshold $\varepsilon ^ *$. What's more, by comparing  Fig.~\ref{fig1new}(a) and (b), one can observe that the advantage of the improved BP algorithm is more significant when the average degree is further increased. As we known, the network is denser when its average degree is increased, in this case, the number of neighbors are comparable to the network size. Approximation in the original BP algorithm becomes inaccurate, leading to worse performance in community detection.

\subsection{CP detection}

The CP structure is different from the community structure, because the nodes in core group not only connect to the nodes in core group but also connect to the nodes in peripheral group, which causes nodal degrees in core group are very large. We first verify the advantages of our improved BP algorithm by considering two real networks.

\begin{figure}
\includegraphics[width=\linewidth]{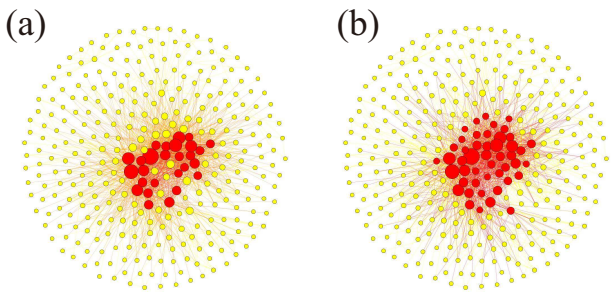}
\caption{(Color online) Detection of CP structure in the USA air network. (a) CP structure is detected by the original BP algorithm, there are 27 nodes in core group. (a) CP structure is detected by the improved BP algorithm and MC sampling, there are 47 nodes in core group. The nodes in core group and peripheral groups are marked by red color and yellow color, respectively.}
\label{fig2}
\end{figure}

The first network is the USA air network with 332 nodes representing the airports and 2126 edges representing the flight routes~\cite{xiang2018unified}. For the original BP algorithm, 27 nodes are assigned to the core group [see the red nodes in Fig.~\ref{fig2}(a)], and the parameters solved by BP original are given as
 \begin{eqnarray}\label{result_BP_air}
{\gamma _{O\_BP}} = \left[ \begin{array}{l}
0.081\\
0.919
\end{array} \right],{P_{O\_BP}} = \left[ {\begin{array}{*{20}{c}}
{0.873}&{0.151}\\
{0.151}&{0.012}
\end{array}} \right].
\end{eqnarray}
However, the improved BP algorithm suggests that there are 47 nodes in the core group [see the red nodes in Fig.~\ref{fig2}(b)], and the parameters in SBM solved by improved BP algorithm are
 \begin{eqnarray}\label{result_new_BP_air}
 {\gamma _{I\_BP}} = \left[ \begin{array}{l}
0.142\\
0.858
\end{array} \right],{P_{I\_BP}} = \left[ {\begin{array}{*{20}{c}}
{0.711}&{0.074}\\
{0.074}&{0.008}
\end{array}} \right].
\end{eqnarray}
We also apply MC sampling to solve the EM algorithm to judge which algorithm is much better. The sampling times are 1000 rounds ($n$ times per round). Surprisingly, the MC sampling and the improved BP algorithm yield the same core nodes and the almost same parameters [see Eq.~(\ref{result_MC})].
\begin{eqnarray}\label{result_MC}
 {\gamma _{MC}} = \left[ \begin{array}{l}
0.142\\
0.858
\end{array} \right],{P_{MC}} = \left[ {\begin{array}{*{20}{c}}
{0.715}&{0.074}\\
{0.074}&{0.008}
\end{array}} \right].
\end{eqnarray}
They are quite different from the original BP algorithm. Therefore, the improved algorithm yields a more accurate solution.

\begin{figure}
\includegraphics[width=\linewidth]{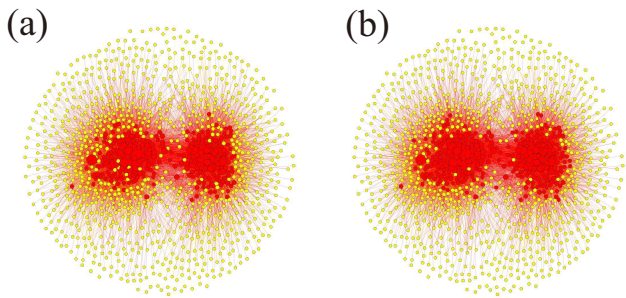}
\caption{(Color online) Detection of CP structure in the Political blogs network. (a) CP structure is detected by the original BP algorithm, there are 294 nodes in core group. (a) CP structure is detected by the improved BP algorithm, there are 336 nodes in core group. The nodes in core group and peripheral groups are marked by red color and yellow color, respectively.}
\label{fig3}
\end{figure}

The second network is the Political blogs network with 1222 nodes and 16714 edges~\cite{ma2016playing}. The original BP algorithm detects 294 core nodes [see the red nodes in Fig.~\ref{fig3}(a)], which is 41 fewer than the 335 nodes solved by the MC sampling. Meanwhile, the parameters solved by the original BP algorithm
 \begin{eqnarray}\label{result_plo_PB}
 {\gamma _{O\_BP}} = \left[ \begin{array}{l}
0.241\\
0.759
\end{array} \right],{P_{O\_BP}} = \left[ {\begin{array}{*{20}{c}}
{0.183}&{0.028}\\
{0.028}&{0.003}
\end{array}} \right]
\end{eqnarray}
 and by the MC sampling
 \begin{eqnarray}\label{result_plo_MC}
 {\gamma _{MC}} = \left[ \begin{array}{l}
0.276\\
0.724
\end{array} \right],{P_{MC}} = \left[ {\begin{array}{*{20}{c}}
{0.161}&{0.023}\\
{0.023}&{0.002}
\end{array}} \right]
\end{eqnarray}
are quite different. The improved BP algorithm detects 336 core nodes [see the red nodes in Fig.~\ref{fig3}(b)], only one more core node is found by the improved BP algorithm. Moreover, the parameters solved by the improved BP algorithm [see  Eq.~(\ref{result_plo_BP1})] are the same as Eq.~(\ref{result_plo_MC}).

 \begin{eqnarray}\label{result_plo_BP1}
 {\gamma _{I\_BP}} = \left[ \begin{array}{l}
0.276\\
0.724
\end{array} \right],{P_{I\_BP}} = \left[ {\begin{array}{*{20}{c}}
{0.161}&{0.023}\\
{0.023}&{0.002}
\end{array}} \right].
\end{eqnarray}

\begin{figure}
\includegraphics[width=\linewidth]{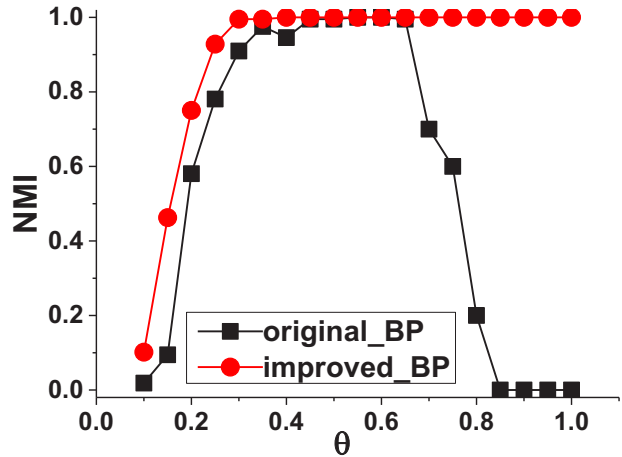}
\caption{(Color online) The value of NMI as a function of the probability of $p_{cc}=\theta$. Parameters: $p_{pp}=0.05$, $p_{cp}=\frac{3}{5}\theta$, the number of nodes in core group and peripheral group are 50 and 150, respectively.  The results are obtained by ensemble averaging over ten independent realizations. }
\label{fig4}
\end{figure}

Then we validate our improved BP algorithm on the synthetic networks with two groups. Given three parameters: the probability of the connection between the core nodes $p_{cc}$, the probability of the connection between the core nodes and the peripheral nodes $p_{cp}$, and the probability of the connection between the peripheral nodes $p_{pp}$, we can generate networks with a CP structure by setting ${p_{cc}} > {p_{cp}} > {p_{pp}}$. In doing so, we set $p_{pp}=0.05$, $p_{cc}=\theta$ and $p_{cp}=\frac{3}{5}\theta$, then a series of synthetic networks can be generated by changing the value of $\theta$. Here we set the network size $n=200$, and the number of core nodes and peripheral nodes are 50 and 150, respectively. From Fig. ~\ref{fig4}, we can see that the improved BP  algorithm is better than the original BP algorithm when the core density is small, i.e., $\theta$ is small. When $\theta$ is gradually increased, both of them can fully detect the CP structure. Nevertheless, with the further increasing $\theta$, the original BP algorithm becomes unstable and cannot detect CP structure any more. The reason is that the core nodes have large degrees when $\theta$ is very large, and the approximation in Eq.~(\ref{approximation}) does not hold any more, leading to the invalidity of the original BP algorithm. However, figure ~\ref{fig4} demonstrates that the improved BP algorithm can solve the problem perfectly.

\section{Conclusions}

In this paper, we have examined the original BP algorithm used for the detection of the meso-scale structures, and found that one approximation in the derivation of the BP algorithm does not hold if some nodal degrees are very large degrees. For example, for the networks with CP structure, the core nodes usually have very large degrees. Therefore, we proposed an improved BP algorithm by avoiding such an approximation. Our experimental results indicates that, even though the modifications are slight, the improved BP algorithm can better detect meso-scale structures without adding any computational complexity, especially for the detection of the CP structure.
 \section*{Acknowledgments}
 This work is supported by National Natural Science Foundation of China (61473001), and partially supported by the Young Talent Funding of Anhui Provincial Universities (gxyqZD2017003).

\begin{thebibliography}{28}%
\makeatletter
\providecommand \@ifxundefined [1]{%
 \@ifx{#1\undefined}
}%
\providecommand \@ifnum [1]{%
 \ifnum #1\expandafter \@firstoftwo
 \else \expandafter \@secondoftwo
 \fi
}%
\providecommand \@ifx [1]{%
 \ifx #1\expandafter \@firstoftwo
 \else \expandafter \@secondoftwo
 \fi
}%
\providecommand \natexlab [1]{#1}%
\providecommand \enquote  [1]{``#1''}%
\providecommand \bibnamefont  [1]{#1}%
\providecommand \bibfnamefont [1]{#1}%
\providecommand \citenamefont [1]{#1}%
\providecommand \href@noop [0]{\@secondoftwo}%
\providecommand \href [0]{\begingroup \@sanitize@url \@href}%
\providecommand \@href[1]{\@@startlink{#1}\@@href}%
\providecommand \@@href[1]{\endgroup#1\@@endlink}%
\providecommand \@sanitize@url [0]{\catcode `\\12\catcode `\$12\catcode
  `\&12\catcode `\#12\catcode `\^12\catcode `\_12\catcode `\%12\relax}%
\providecommand \@@startlink[1]{}%
\providecommand \@@endlink[0]{}%
\providecommand \url  [0]{\begingroup\@sanitize@url \@url }%
\providecommand \@url [1]{\endgroup\@href {#1}{\urlprefix }}%
\providecommand \urlprefix  [0]{URL }%
\providecommand \Eprint [0]{\href }%
\providecommand \doibase [0]{http://dx.doi.org/}%
\providecommand \selectlanguage [0]{\@gobble}%
\providecommand \bibinfo  [0]{\@secondoftwo}%
\providecommand \bibfield  [0]{\@secondoftwo}%
\providecommand \translation [1]{[#1]}%
\providecommand \BibitemOpen [0]{}%
\providecommand \bibitemStop [0]{}%
\providecommand \bibitemNoStop [0]{.\EOS\space}%
\providecommand \EOS [0]{\spacefactor3000\relax}%
\providecommand \BibitemShut  [1]{\csname bibitem#1\endcsname}%
\let\auto@bib@innerbib\@empty
\bibitem [{\citenamefont {Fortunato}(2010)}]{fortunato2010community}%
  \BibitemOpen
  \bibfield  {author} {\bibinfo {author} {\bibfnamefont {S.}~\bibnamefont
  {Fortunato}},\ }\href@noop {} {\bibfield  {journal} {\bibinfo  {journal}
  {Physics Reports}\ }\textbf {\bibinfo {volume} {486}},\ \bibinfo {pages} {75}
  (\bibinfo {year} {2010})}\BibitemShut {NoStop}%
\bibitem [{\citenamefont {Newman}(2006)}]{newman2006modularity}%
  \BibitemOpen
  \bibfield  {author} {\bibinfo {author} {\bibfnamefont {M.~E.}\ \bibnamefont
  {Newman}},\ }\href@noop {} {\bibfield  {journal} {\bibinfo  {journal}
  {Proceedings of the National Academy of Sciences}\ }\textbf {\bibinfo
  {volume} {103}},\ \bibinfo {pages} {8577} (\bibinfo {year}
  {2006})}\BibitemShut {NoStop}%
\bibitem [{\citenamefont {White}\ and\ \citenamefont
  {Smyth}(2005)}]{white2005spectral}%
  \BibitemOpen
  \bibfield  {author} {\bibinfo {author} {\bibfnamefont {S.}~\bibnamefont
  {White}}\ and\ \bibinfo {author} {\bibfnamefont {P.}~\bibnamefont {Smyth}},\
  }in\ \href@noop {} {\emph {\bibinfo {booktitle} {Proceedings of the 2005 SIAM
  international conference on data mining}}}\ (\bibinfo {organization} {SIAM},\
  \bibinfo {year} {2005})\ pp.\ \bibinfo {pages} {274--285}\BibitemShut
  {NoStop}%
\bibitem [{\citenamefont {Lancichinetti}\ \emph {et~al.}(2009)\citenamefont
  {Lancichinetti}, \citenamefont {Fortunato},\ and\ \citenamefont
  {Kert{\'e}sz}}]{lancichinetti2009detecting}%
  \BibitemOpen
  \bibfield  {author} {\bibinfo {author} {\bibfnamefont {A.}~\bibnamefont
  {Lancichinetti}}, \bibinfo {author} {\bibfnamefont {S.}~\bibnamefont
  {Fortunato}}, \ and\ \bibinfo {author} {\bibfnamefont {J.}~\bibnamefont
  {Kert{\'e}sz}},\ }\href@noop {} {\bibfield  {journal} {\bibinfo  {journal}
  {New Journal of Physics}\ }\textbf {\bibinfo {volume} {11}},\ \bibinfo
  {pages} {033015} (\bibinfo {year} {2009})}\BibitemShut {NoStop}%
\bibitem [{\citenamefont {Girvan}\ and\ \citenamefont
  {Newman}(2002)}]{girvan2002community}%
  \BibitemOpen
  \bibfield  {author} {\bibinfo {author} {\bibfnamefont {M.}~\bibnamefont
  {Girvan}}\ and\ \bibinfo {author} {\bibfnamefont {M.~E.}\ \bibnamefont
  {Newman}},\ }\href@noop {} {\bibfield  {journal} {\bibinfo  {journal}
  {Proceedings of the National Academy of Sciences}\ }\textbf {\bibinfo
  {volume} {99}},\ \bibinfo {pages} {7821} (\bibinfo {year}
  {2002})}\BibitemShut {NoStop}%
\bibitem [{\citenamefont {Yang}\ and\ \citenamefont
  {Leskovec}(2013)}]{yang2013overlapping}%
  \BibitemOpen
  \bibfield  {author} {\bibinfo {author} {\bibfnamefont {J.}~\bibnamefont
  {Yang}}\ and\ \bibinfo {author} {\bibfnamefont {J.}~\bibnamefont
  {Leskovec}},\ }in\ \href@noop {} {\emph {\bibinfo {booktitle} {Proceedings of
  the sixth ACM international conference on Web search and data mining}}}\
  (\bibinfo {organization} {ACM},\ \bibinfo {year} {2013})\ pp.\ \bibinfo
  {pages} {587--596}\BibitemShut {NoStop}%
\bibitem [{\citenamefont {Palla}\ \emph {et~al.}(2005)\citenamefont {Palla},
  \citenamefont {Der{\'e}nyi}, \citenamefont {Farkas},\ and\ \citenamefont
  {Vicsek}}]{palla2005uncovering}%
  \BibitemOpen
  \bibfield  {author} {\bibinfo {author} {\bibfnamefont {G.}~\bibnamefont
  {Palla}}, \bibinfo {author} {\bibfnamefont {I.}~\bibnamefont {Der{\'e}nyi}},
  \bibinfo {author} {\bibfnamefont {I.}~\bibnamefont {Farkas}}, \ and\ \bibinfo
  {author} {\bibfnamefont {T.}~\bibnamefont {Vicsek}},\ }\href@noop {}
  {\bibfield  {journal} {\bibinfo  {journal} {Nature}\ }\textbf {\bibinfo
  {volume} {435}},\ \bibinfo {pages} {814} (\bibinfo {year}
  {2005})}\BibitemShut {NoStop}%
\bibitem [{\citenamefont {White}\ \emph {et~al.}(1976)\citenamefont {White},
  \citenamefont {Boorman},\ and\ \citenamefont {Breiger}}]{white1976social}%
  \BibitemOpen
  \bibfield  {author} {\bibinfo {author} {\bibfnamefont {H.~C.}\ \bibnamefont
  {White}}, \bibinfo {author} {\bibfnamefont {S.~A.}\ \bibnamefont {Boorman}},
  \ and\ \bibinfo {author} {\bibfnamefont {R.~L.}\ \bibnamefont {Breiger}},\
  }\href@noop {} {\bibfield  {journal} {\bibinfo  {journal} {American Journal
  of Sociology}\ ,\ \bibinfo {pages} {730}} (\bibinfo {year}
  {1976})}\BibitemShut {NoStop}%
\bibitem [{\citenamefont {Borgatti}\ and\ \citenamefont
  {Everett}(2000)}]{borgatti2000models}%
  \BibitemOpen
  \bibfield  {author} {\bibinfo {author} {\bibfnamefont {S.~P.}\ \bibnamefont
  {Borgatti}}\ and\ \bibinfo {author} {\bibfnamefont {M.~G.}\ \bibnamefont
  {Everett}},\ }\href@noop {} {\bibfield  {journal} {\bibinfo  {journal}
  {Social networks}\ }\textbf {\bibinfo {volume} {21}},\ \bibinfo {pages} {375}
  (\bibinfo {year} {2000})}\BibitemShut {NoStop}%
\bibitem [{\citenamefont {Rombach}\ \emph {et~al.}(2017)\citenamefont
  {Rombach}, \citenamefont {Porter}, \citenamefont {Fowler},\ and\
  \citenamefont {Mucha}}]{rombach2017core}%
  \BibitemOpen
  \bibfield  {author} {\bibinfo {author} {\bibfnamefont {P.}~\bibnamefont
  {Rombach}}, \bibinfo {author} {\bibfnamefont {M.~A.}\ \bibnamefont {Porter}},
  \bibinfo {author} {\bibfnamefont {J.~H.}\ \bibnamefont {Fowler}}, \ and\
  \bibinfo {author} {\bibfnamefont {P.~J.}\ \bibnamefont {Mucha}},\ }\href@noop
  {} {\bibfield  {journal} {\bibinfo  {journal} {SIAM Review}\ }\textbf
  {\bibinfo {volume} {59}},\ \bibinfo {pages} {619} (\bibinfo {year}
  {2017})}\BibitemShut {NoStop}%
\bibitem [{\citenamefont {Verma}\ \emph {et~al.}(2016)\citenamefont {Verma},
  \citenamefont {Russmann}, \citenamefont {Ara{\'u}jo}, \citenamefont
  {Nagler},\ and\ \citenamefont {Herrmann}}]{verma2016emergence}%
  \BibitemOpen
  \bibfield  {author} {\bibinfo {author} {\bibfnamefont {T.}~\bibnamefont
  {Verma}}, \bibinfo {author} {\bibfnamefont {F.}~\bibnamefont {Russmann}},
  \bibinfo {author} {\bibfnamefont {N.}~\bibnamefont {Ara{\'u}jo}}, \bibinfo
  {author} {\bibfnamefont {J.}~\bibnamefont {Nagler}}, \ and\ \bibinfo {author}
  {\bibfnamefont {H.}~\bibnamefont {Herrmann}},\ }\href@noop {} {\bibfield
  {journal} {\bibinfo  {journal} {Nature Communications}\ }\textbf {\bibinfo
  {volume} {7}},\ \bibinfo {pages} {10441} (\bibinfo {year}
  {2016})}\BibitemShut {NoStop}%
\bibitem [{\citenamefont {Chen}\ \emph {et~al.}(2018)\citenamefont {Chen},
  \citenamefont {Zhang},\ and\ \citenamefont {Shen}}]{chen2018double}%
  \BibitemOpen
  \bibfield  {author} {\bibinfo {author} {\bibfnamefont {H.}~\bibnamefont
  {Chen}}, \bibinfo {author} {\bibfnamefont {H.}~\bibnamefont {Zhang}}, \ and\
  \bibinfo {author} {\bibfnamefont {C.}~\bibnamefont {Shen}},\ }\href@noop {}
  {\bibfield  {journal} {\bibinfo  {journal} {Journal of Statistical Mechanics:
  Theory and Experiment}\ }\textbf {\bibinfo {volume} {2018}},\ \bibinfo
  {pages} {063402} (\bibinfo {year} {2018})}\BibitemShut {NoStop}%
\bibitem [{\citenamefont {Kojaku}\ and\ \citenamefont
  {Masuda}(2017)}]{kojaku2017finding}%
  \BibitemOpen
  \bibfield  {author} {\bibinfo {author} {\bibfnamefont {S.}~\bibnamefont
  {Kojaku}}\ and\ \bibinfo {author} {\bibfnamefont {N.}~\bibnamefont
  {Masuda}},\ }\href@noop {} {\bibfield  {journal} {\bibinfo  {journal}
  {Physical Review E}\ }\textbf {\bibinfo {volume} {96}},\ \bibinfo {pages}
  {052313} (\bibinfo {year} {2017})}\BibitemShut {NoStop}%
\bibitem [{\citenamefont {Xiang}\ \emph {et~al.}(2018)\citenamefont {Xiang},
  \citenamefont {Bao}, \citenamefont {Ma}, \citenamefont {Zhang}, \citenamefont
  {Chen},\ and\ \citenamefont {Zhang}}]{xiang2018unified}%
  \BibitemOpen
  \bibfield  {author} {\bibinfo {author} {\bibfnamefont {B.-B.}\ \bibnamefont
  {Xiang}}, \bibinfo {author} {\bibfnamefont {Z.-K.}\ \bibnamefont {Bao}},
  \bibinfo {author} {\bibfnamefont {C.}~\bibnamefont {Ma}}, \bibinfo {author}
  {\bibfnamefont {X.}~\bibnamefont {Zhang}}, \bibinfo {author} {\bibfnamefont
  {H.-S.}\ \bibnamefont {Chen}}, \ and\ \bibinfo {author} {\bibfnamefont
  {H.-F.}\ \bibnamefont {Zhang}},\ }\href@noop {} {\bibfield  {journal}
  {\bibinfo  {journal} {Chaos: An Interdisciplinary Journal of Nonlinear
  Science}\ }\textbf {\bibinfo {volume} {28}},\ \bibinfo {pages} {013122}
  (\bibinfo {year} {2018})}\BibitemShut {NoStop}%
\bibitem [{\citenamefont {Della~Rossa}\ \emph {et~al.}(2013)\citenamefont
  {Della~Rossa}, \citenamefont {Dercole},\ and\ \citenamefont
  {Piccardi}}]{della2013profiling}%
  \BibitemOpen
  \bibfield  {author} {\bibinfo {author} {\bibfnamefont {F.}~\bibnamefont
  {Della~Rossa}}, \bibinfo {author} {\bibfnamefont {F.}~\bibnamefont
  {Dercole}}, \ and\ \bibinfo {author} {\bibfnamefont {C.}~\bibnamefont
  {Piccardi}},\ }\href@noop {} {\bibfield  {journal} {\bibinfo  {journal}
  {Scientific Reports}\ }\textbf {\bibinfo {volume} {3}},\ \bibinfo {pages}
  {1467} (\bibinfo {year} {2013})}\BibitemShut {NoStop}%
\bibitem [{\citenamefont {Ma}\ \emph {et~al.}(2018)\citenamefont {Ma},
  \citenamefont {Xiang}, \citenamefont {Chen}, \citenamefont {Small},\ and\
  \citenamefont {Zhang}}]{ma2018detection}%
  \BibitemOpen
  \bibfield  {author} {\bibinfo {author} {\bibfnamefont {C.}~\bibnamefont
  {Ma}}, \bibinfo {author} {\bibfnamefont {B.-B.}\ \bibnamefont {Xiang}},
  \bibinfo {author} {\bibfnamefont {H.-S.}\ \bibnamefont {Chen}}, \bibinfo
  {author} {\bibfnamefont {M.}~\bibnamefont {Small}}, \ and\ \bibinfo {author}
  {\bibfnamefont {H.-F.}\ \bibnamefont {Zhang}},\ }\href@noop {} {\bibfield
  {journal} {\bibinfo  {journal} {Chaos: An Interdisciplinary Journal of
  Nonlinear Science}\ }\textbf {\bibinfo {volume} {28}},\ \bibinfo {pages}
  {053121} (\bibinfo {year} {2018})}\BibitemShut {NoStop}%
\bibitem [{\citenamefont {Holme}(2005)}]{holme2005core}%
  \BibitemOpen
  \bibfield  {author} {\bibinfo {author} {\bibfnamefont {P.}~\bibnamefont
  {Holme}},\ }\href@noop {} {\bibfield  {journal} {\bibinfo  {journal}
  {Physical Review E}\ }\textbf {\bibinfo {volume} {72}},\ \bibinfo {pages}
  {046111} (\bibinfo {year} {2005})}\BibitemShut {NoStop}%
\bibitem [{\citenamefont {Kojaku}\ and\ \citenamefont
  {Masuda}(2018)}]{kojaku2018core}%
  \BibitemOpen
  \bibfield  {author} {\bibinfo {author} {\bibfnamefont {S.}~\bibnamefont
  {Kojaku}}\ and\ \bibinfo {author} {\bibfnamefont {N.}~\bibnamefont
  {Masuda}},\ }\href@noop {} {\bibfield  {journal} {\bibinfo  {journal} {New
  Journal of Physics}\ }\textbf {\bibinfo {volume} {20}},\ \bibinfo {pages}
  {043012} (\bibinfo {year} {2018})}\BibitemShut {NoStop}%
\bibitem [{\citenamefont {Decelle}\ \emph
  {et~al.}(2011{\natexlab{a}})\citenamefont {Decelle}, \citenamefont
  {Krzakala}, \citenamefont {Moore},\ and\ \citenamefont
  {Zdeborov{\'a}}}]{decelle2011inference}%
  \BibitemOpen
  \bibfield  {author} {\bibinfo {author} {\bibfnamefont {A.}~\bibnamefont
  {Decelle}}, \bibinfo {author} {\bibfnamefont {F.}~\bibnamefont {Krzakala}},
  \bibinfo {author} {\bibfnamefont {C.}~\bibnamefont {Moore}}, \ and\ \bibinfo
  {author} {\bibfnamefont {L.}~\bibnamefont {Zdeborov{\'a}}},\ }\href@noop {}
  {\bibfield  {journal} {\bibinfo  {journal} {Physical Review Letters}\
  }\textbf {\bibinfo {volume} {107}},\ \bibinfo {pages} {065701} (\bibinfo
  {year} {2011}{\natexlab{a}})}\BibitemShut {NoStop}%
\bibitem [{\citenamefont {Decelle}\ \emph
  {et~al.}(2011{\natexlab{b}})\citenamefont {Decelle}, \citenamefont
  {Krzakala}, \citenamefont {Moore},\ and\ \citenamefont
  {Zdeborov{\'a}}}]{decelle2011asymptotic}%
  \BibitemOpen
  \bibfield  {author} {\bibinfo {author} {\bibfnamefont {A.}~\bibnamefont
  {Decelle}}, \bibinfo {author} {\bibfnamefont {F.}~\bibnamefont {Krzakala}},
  \bibinfo {author} {\bibfnamefont {C.}~\bibnamefont {Moore}}, \ and\ \bibinfo
  {author} {\bibfnamefont {L.}~\bibnamefont {Zdeborov{\'a}}},\ }\href@noop {}
  {\bibfield  {journal} {\bibinfo  {journal} {Physical Review E}\ }\textbf
  {\bibinfo {volume} {84}},\ \bibinfo {pages} {066106} (\bibinfo {year}
  {2011}{\natexlab{b}})}\BibitemShut {NoStop}%
\bibitem [{\citenamefont {Zhang}\ and\ \citenamefont
  {Moore}(2014)}]{zhang2014scalable}%
  \BibitemOpen
  \bibfield  {author} {\bibinfo {author} {\bibfnamefont {P.}~\bibnamefont
  {Zhang}}\ and\ \bibinfo {author} {\bibfnamefont {C.}~\bibnamefont {Moore}},\
  }\href@noop {} {\bibfield  {journal} {\bibinfo  {journal} {Proceedings of the
  National Academy of Sciences}\ }\textbf {\bibinfo {volume} {111}},\ \bibinfo
  {pages} {18144} (\bibinfo {year} {2014})}\BibitemShut {NoStop}%
\bibitem [{\citenamefont {Karrer}\ and\ \citenamefont
  {Newman}(2011)}]{karrer2011stochastic}%
  \BibitemOpen
  \bibfield  {author} {\bibinfo {author} {\bibfnamefont {B.}~\bibnamefont
  {Karrer}}\ and\ \bibinfo {author} {\bibfnamefont {M.~E.}\ \bibnamefont
  {Newman}},\ }\href@noop {} {\bibfield  {journal} {\bibinfo  {journal}
  {Physical Review E}\ }\textbf {\bibinfo {volume} {83}},\ \bibinfo {pages}
  {016107} (\bibinfo {year} {2011})}\BibitemShut {NoStop}%
\bibitem [{\citenamefont {Zhang}\ \emph {et~al.}(2015)\citenamefont {Zhang},
  \citenamefont {Martin},\ and\ \citenamefont
  {Newman}}]{zhang2015identification}%
  \BibitemOpen
  \bibfield  {author} {\bibinfo {author} {\bibfnamefont {X.}~\bibnamefont
  {Zhang}}, \bibinfo {author} {\bibfnamefont {T.}~\bibnamefont {Martin}}, \
  and\ \bibinfo {author} {\bibfnamefont {M.~E.}\ \bibnamefont {Newman}},\
  }\href@noop {} {\bibfield  {journal} {\bibinfo  {journal} {Physical Review
  E}\ }\textbf {\bibinfo {volume} {91}},\ \bibinfo {pages} {032803} (\bibinfo
  {year} {2015})}\BibitemShut {NoStop}%
\bibitem [{\citenamefont {Dempster}\ \emph {et~al.}(1977)\citenamefont
  {Dempster}, \citenamefont {Laird},\ and\ \citenamefont {Rubin}}]{EM1977}%
  \BibitemOpen
  \bibfield  {author} {\bibinfo {author} {\bibfnamefont {P.}~\bibnamefont
  {Dempster}, \bibfnamefont {A.}}, \bibinfo {author} {\bibfnamefont {N.~M.}\
  \bibnamefont {Laird}}, \ and\ \bibinfo {author} {\bibfnamefont
  {B.}~\bibnamefont {Rubin}, \bibfnamefont {D.}},\ }\href@noop {} {\bibfield
  {journal} {\bibinfo  {journal} {Journal of the Royal Statistical Society.
  Series B (Methodological)}\ }\textbf {\bibinfo {volume} {39}},\ \bibinfo
  {pages} {1} (\bibinfo {year} {1977})}\BibitemShut {NoStop}%
\bibitem [{\citenamefont {Pearl}(1988)}]{lamperti2012stochastic}%
  \BibitemOpen
  \bibfield  {author} {\bibinfo {author} {\bibfnamefont {J.}~\bibnamefont
  {Pearl}},\ }\href@noop {} {\emph {\bibinfo {title} {Probabilistic reasoning
  in intelligent systems : networks of plausible inference}}}\ (\bibinfo
  {publisher} {Morgan Kaufmann},\ \bibinfo {year} {1988})\BibitemShut {NoStop}%
\bibitem [{\citenamefont {Shi}\ \emph {et~al.}(2018)\citenamefont {Shi},
  \citenamefont {Liu},\ and\ \citenamefont {Zhang}}]{shi2018weighted}%
  \BibitemOpen
  \bibfield  {author} {\bibinfo {author} {\bibfnamefont {C.}~\bibnamefont
  {Shi}}, \bibinfo {author} {\bibfnamefont {Y.}~\bibnamefont {Liu}}, \ and\
  \bibinfo {author} {\bibfnamefont {P.}~\bibnamefont {Zhang}},\ }\href@noop {}
  {\bibfield  {journal} {\bibinfo  {journal} {Journal of Statistical Mechanics:
  Theory and Experiment}\ }\textbf {\bibinfo {volume} {2018}},\ \bibinfo
  {pages} {033405} (\bibinfo {year} {2018})}\BibitemShut {NoStop}%
\bibitem [{\citenamefont {Pizzuti}(2012)}]{pizzuti2012multiobjective}%
  \BibitemOpen
  \bibfield  {author} {\bibinfo {author} {\bibfnamefont {C.}~\bibnamefont
  {Pizzuti}},\ }\href@noop {} {\bibfield  {journal} {\bibinfo  {journal} {IEEE
  Transactions on Evolutionary Computation}\ }\textbf {\bibinfo {volume}
  {16}},\ \bibinfo {pages} {418} (\bibinfo {year} {2012})}\BibitemShut
  {NoStop}%
\bibitem [{\citenamefont {Ma}\ \emph {et~al.}(2016)\citenamefont {Ma},
  \citenamefont {Zhou},\ and\ \citenamefont {Zhang}}]{ma2016playing}%
  \BibitemOpen
  \bibfield  {author} {\bibinfo {author} {\bibfnamefont {C.}~\bibnamefont
  {Ma}}, \bibinfo {author} {\bibfnamefont {T.}~\bibnamefont {Zhou}}, \ and\
  \bibinfo {author} {\bibfnamefont {H.-F.}\ \bibnamefont {Zhang}},\ }\href@noop
  {} {\bibfield  {journal} {\bibinfo  {journal} {Scientific Reports}\ }\textbf
  {\bibinfo {volume} {6}},\ \bibinfo {pages} {30098} (\bibinfo {year}
  {2016})}\BibitemShut {NoStop}%
\end{thebibliography}

%

\end{document}